\documentclass[prl,twocolumn,showpacs,nofootinbib]{revtex4}
\usepackage{amsmath}
\usepackage{amssymb}
\usepackage{amsthm}
\usepackage{ifthen}
\usepackage{enumerate}

\newboolean{showdetails}
\setboolean{showdetails}{false}

\newtheorem{thm}{Theorem}

\newcommand{\F}{\mathcal F}
\newcommand{\Z}{\mathbb Z}
\newcommand{\ham}{\mathcal H}
\newcommand{\G}{\widetilde{G}_{\Lambda}}

\newcommand{\gf}{G^\delta_{\Lambda,\Gamma}}

\newcommand{\gr}{g_{\Lambda,\Gamma}^{\delta}}

\newcommand{\lop}{\kappa}

\newcommand{\Tr}{\textup{Tr }}

\newcommand{\E}[1]{\textup{Av} \left[#1\right]}

\newcommand{\condExp}[2]{\ensuremath{\textup{Av}\left[ #1 \middle| #2 \right] }}

\newcommand{\Sys}{QRFIM}
\newcommand{\co}{\ensuremath{^c}}
\newcommand{\Zd}{\ensuremath{\mathbb{Z}^d}}

\newcommand{\m}[1]{\ensuremath{m_{#1}(T,h,\epsilon)}}

\begin{document}
\title{Rounding of First Order Transitions in Low-Dimensional Quantum Systems with Quenched Disorder} \author{Rafael L. Greenblatt}
\email{rafaelgr@physics.rutgers.edu}
\affiliation{Department of Physics and Astronomy, Rutgers University, Piscataway NJ 08854-8019, USA} \author{Michael Aizenman}
\email{aizenman@princeton.edu}
\affiliation{Departments of Physics and Mathematics, Princeton University, Princeton NJ 08544, USA} \author{Joel L. Lebowitz}
\email{lebowitz@math.rutgers.edu}
\affiliation{Departments of Mathematics and Physics, Rutgers University, Piscataway NJ 08854-8019, USA} \date{\today}

\begin{abstract}
We prove that the addition of an arbitrarily small random perturbation to a quantum spin system rounds a first order phase transition in the conjugate order parameter in $d \leq 2$ dimensions, or for cases involving the breaking of a continuous symmetry in $d \leq 4$. This establishes rigorously for quantum systems the existence of the Imry-Ma phenomenon
which for classical systems was proven by Aizenman and Wehr.
\end{abstract}
\maketitle

A first order phase transition, in Ehrenfest's  terminology, is one associated with a discontinuity in the density of an extensive quantity.   In  thermodynamic terms this corresponds to a discontinuity in the derivative of the free energy with respect to one of the parameters in the Hamiltonian, more specifically the one conjugate  to the order parameter, e.g. the magnetic field in a ferromagnetic spin system.
In what is known as the Imry-Ma phenomenon~\cite{YM,AYM}, any such discontinuity is rounded off in low dimensions when the Hamiltonian of a homogeneous system is modified through the  incorporation of an arbitrarily weak random term, corresponding to quenched local disorder,  in the field conjugate to the order parameter.

This phenomenon has been rigorously established for classical systems~\cite{AW.PRL,AW}, where it occurs  in  dimensions $d\le 2$, and $d\le 4$ when  the discontinuity is associated with the breaking of a continuous symmetry.   In this letter we prove analogous results for quantum systems at both positive and zero temperatures (ground states).

The existence of this effect was first argued for random fields by Imry and Ma  on the basis of a heuristic analysis of free energy fluctuations.  While the sufficiency of Imry and Ma's reasoning was called into question, the predicted phenomenon was established rigorously through a number of works~\cite{FFS,Imbrie.PRL,Imbrie.CMP,BK.PRL,BK.CMP,AW.PRL,AW}.  The statement was further extended to different disorder types by Hui and Berker~\cite{HuiBerker.PRL,Berker.PRB.90}.


The general existence of the Imry-Ma phenomenon in quantum systems was not addressed by these rigorous analysis, and in particular the Aizenman-Wehr~\cite{AW,AW.PRL} proof of the rounding effect applies only for classical systems.   However, as  stressed in~\cite{Goswami},  establishing whether  the Imry-Ma phenomenon extends to first order quantum phase transitions (QPT$_1$)  is an important open problem.   The results presented here answer this question.
We find that the critical dimensions for the phenomenon for  quantum systems are the same as for classical systems, including at zero temperature.

%

We consider spin systems on the $d$-dimensional lattice $\mathbb{Z}^d$, where the configuration at each site is described by a finite-dimensional Hilbert space, with a Hamiltonian of the form
\begin{equation}
\label{Hamiltonian}
\mathcal{H} = \mathcal{H}_0 - \sum_x \left( h + \epsilon \eta_{x} \right) \lop_{x} \end{equation}
where $\{\lop_x\}$ are translates of some local operator $\lop_0$,
 and $h$ and $\epsilon$ are real parameters.  The quenched disorder is represented by $\{\eta_x\}$, a family of independent, identically distributed random variables.  $\mathcal{H}_0$ may be translation invariant and nonrandom, or it can include additional random terms (although we will not discuss the latter case, our results hold there also).
For convenience we will assume that $\|\lop_x\| = 1$, which can be arranged by rescaling $h$ and $\epsilon$.
We will refer to the $\eta$s as random fields, although in general they may also be associated with some other parameters, e.g.  random bond strengths.

An example of a system of this type (with $\lop_x=\sigma_x^{(3)}$) is the ferromagnetic transverse-field Ising model with a random longitudinal field~\cite{Senthil} (henceforth \Sys), with
 \begin{equation} \label{thisHam}
\mathcal{H} = -\sum J_{x-y} \sigma_x^{(3)} \sigma_y^{(3)} - \sum \left[ \lambda \, \sigma_x^{(1)} + (h+ \epsilon \eta_x) \, \sigma_x^{(3)} \right] \end{equation} where $\sigma_x^{(i)}$ $(i=1,2,3)$ are single-site Pauli matrices, and $J_{x-y} > 0$.
The \Sys~has recently been studied as a model for the behavior of ${\rm LiHo_xY_{1-x}F_4}$ with $x>0.5$ in a strong transverse magnetic field~\cite{Schechter,SchSt.PRB}.

We will examine phase transitions where the order parameter is the volume average of the expectation value of $\lop_x $ with respect to an equilibrium state (that is, a state satisfying the Kubo-Martin-Schwinger condition~\cite{Ruelle}), and show that this quantity cannot be discontinuous in $h$ for low-dimensional systems.  As is well known,
this order parameter is related to the directional derivatives $(\pm)$ of the \emph{free energy density},
\begin{equation}
\m{\pm} \ := \ -\frac{\partial} {\partial h \pm} \F(T, h, \epsilon) \, \label{mDef}
\end{equation}
where, as usual,  at positive temperatures
\begin{equation} \label{Z}
\F(T,h, \epsilon) = \ \lim_{\Gamma \nearrow \Zd}
\frac{-1}{\beta |\Gamma|} \log \Tr e^{-\beta H_{\Gamma}} \, \end{equation}
(with $\beta := 1/k_B T$), and $\F(0,h,\epsilon)$ is the corresponding limit of the ground state energy.
Here $H_{\Gamma}$ is the Hamiltonian of the system restricted to the finite box $\Gamma \subset \Zd$, and $ |\Gamma|$ is the number of sites in that box. It is known under the assumptions enumerated below that for almost all $\eta$ this limit exists and is given by a non-random function of the parameters \ifthenelse{\boolean{showdetails}}{(see Appendix~\ref{A_TL})}{(see, e.g.~\cite{AW,vuillermot1977tqr})}, which does not depend on the boundary conditions.
By general arguments which are valid for both classical and quantum systems, $\F$ is convex in $h$; therefore the directional derivatives exist, and are equal for all but countably many values of $h$ \cite{Ruelle}.

For typical realizations of the random field, the interval $[\m{-}, \m{+}]$ provides the asymptotic range of values of the order parameter for any sequence of finite volume Gibbs states
 or ground states
 \ifthenelse{\boolean{showdetails}}{}{(the argument is similar to that found in \cite{AW} for classical systems)}.  At a first order phase transition $m_- < m_+$, and there are then at least two distinct infinite volume equilibrium states~\cite{Ruelle} with different values of the order parameter.
In the \Sys~the $m_+$ and the $m_-$states  can be obtained through the $+$ or $-$ boundary conditions (i.e. the spins $\sigma^{(3)}_x$ are replaced by $\pm 1$ for all $x \notin \Gamma$).  In general, such states  are  obtained by  adding $\pm \delta$ to the uniform field $h$ and letting  $\delta \to 0$ \emph{after} taking the infinite volume limit.

Our discussion is restricted to systems satisfying:
\begin{enumerate}[A.]
\item \label{finiteRange}
The interactions are \emph{short range}, in the sense that
for any finite box $\Lambda \in \Zd$ the Hamiltonian may be decomposed as: $\ham = H_\Lambda+V_\Lambda+H_{\Lambda\co}$, with $H_\Lambda$ acting only in $\Lambda$,  $H_{\Lambda\co} $  only in the complement  $\Lambda\co$, and $V_\Lambda$ of norm bounded by the size of the boundary:
 \begin{equation}
\| V_\Lambda \| \leq C | \partial \Lambda |  \label{finiteRangeIneq} \, .
\end{equation}
\item \label{fluct} The variables $\eta_x$ have an \emph{absolutely continuous}  distribution with respect to the Lebesgue measure  (i.e. one  with a probability density with no delta functions), and a  finite  \emph{ $r$th moment}, for some $r>2$.
 \end{enumerate}

Our main results are summarized in the following two statements.  The first applies regardless
of whether the order parameter is related to any symmetry breaking.

\begin{thm}\label{bigThm}
In dimensions $d\le 2$, any system of the form of \eqref{Hamiltonian} satisfying the above assumptions has $\m{+}=\m{-}$ for all $h$, and $T\ge 0$, provided $\epsilon \neq 0$.    \end{thm}

The next result is formulated for situations where the first order phase transition would represent continuous symmetry breaking.  An example is the  $O(N)$ model
with
\begin{equation} \label{HeisHam}
 \mathcal{H}_0 \ =\ -  \sum J_{x-y} \vec{\sigma}_x \cdot \vec{\sigma}_y
\end{equation}
where $\vec{\sigma}$ are the usual quantum spin operators.    More generally, $ \mathcal{H}_0 $ is assumed to be a sum of finite range terms which are invariant under the global action of the rotation group $SO(N)$, and $\vec{\sigma}_x$ is a collection of operators of norm
one which transform as  the components of a vector under rotations.
With the random terms the Hamiltonian is
\begin{equation}
\mathcal{H} =  \mathcal{H}_0  - \sum (\vec{h} + \epsilon \vec{\eta}_x) \cdot \vec{\sigma}_x. \label{ONHam} \end{equation}

\begin{thm}\label{bigThm2}
For the $SO(N)$-symmetric system described above, with the random fields $\vec{\eta}_x$ having a rotation-invariant distribution, the free energy
is continuously differentiable in $\vec{h}$ at $\vec{h}=0$ whenever $\epsilon \neq 0$, $d \leq 4$, and $N \geq 2$.
\end{thm}

Before describing the proof, let us comment on   the implications of the statements, and their limitations.
\newcounter{ccount}
\begin{list}{\arabic{ccount}. }{\usecounter{ccount} \setlength{\itemindent}{\leftmargin} \setlength{\listparindent}{\parindent} \setlength{\leftmargin}{0pt}}
\item While the statements establish uniqueness of the expectation value of the bulk averages of the observables $\kappa_x$, or $\vec{\sigma}_x$ (in Theorem 2), they do not rule out the possibility of the coexistence of a number of equilibrium states, which differ from each other in some other way than the mean density of $\kappa$, which they share.
More can be said  for models for which it is known by other means   that non-uniquess of state is possible only if there is long range order in $\kappa$.   (Such is the case for  QRFIM,  through its relation to the classical ferromagnetic Ising model in $d+1$ dimensions~\cite{campanino1991lgs}.)

\item The results address only the discontinuity,  or symmetry breaking (as in the QRFIM), but they leave room for other phase transitions, or singular dependence on $h$.    For instance, for  the Ashkin-Teller spin chain for which Goswami et. al.~\cite{Goswami}  report finding the Imry-Ma phenomenon in some range of the parameters but not elsewhere,  the results presented here rule out the persistence of a first-order transition between the paramagnetic and Baxter phases in the full range in the model's parameters.   However, they do not rule out the possibility of  other phase transitions.

\item Randomness which does not couple to the order parameter of the transition need not cause a rounding effect.  For example, in the transverse-field Ising model in a random \emph{transverse} field, where the random field $\eta_x$ couples to $\sigma^{(1)}$, ferromagnetic ordering is known to persist~\cite{Fisher,campanino1991lgs}.  Presumably the same is true for the Baxter phase of the Ashkin-Teller model.  It was  even suggested that there are systems in which the introduction of randomness of this sort may even induce long range order which would not otherwise be present~\cite{Wehr.PRB,Wehr.PRL}, and our results do not contradict this.  In addition we can draw no conclusions about quasi-long-range order, that is power law decay of correlations, including of $\kappa_x$.

\end{list}

Other comments, on the technical assumptions under which the statements hold, are found after the proofs. \\

The proofs of Theorems~\ref{bigThm} and~\ref{bigThm2} are based on the analysis of the differences, between the $m_+$ and the $m_-$-states, in the free energy (at $T=0$, ground state energy) which can be ascribed   to the random field within a finite region $\Lambda$ of diameter $L$.
Putting momentarily aside  the question of existence of limits, a relevant quantity could be provided by:
\begin{equation} \label{gLim}
\G(\eta_\Lambda) := \lim_{\delta \to 0}\lim_{\Gamma \to \Z^d} \condExp{\gf}{\eta_\Lambda} -\E{\gf}
\end{equation}
where
$\gf$ is the difference of free energies
\begin{equation}
\gf(\eta):= \frac{1}{2} \left( F^{\eta,h+\delta}_{\Gamma} - F^{\eta^{(\Lambda)},h+\delta}_{\Gamma}-F^{\eta,h-\delta}_{\Gamma} + F^{\eta^{(\Lambda)},h-\delta}_{\Gamma}\right), \label{gfDef}
\end{equation}
with
\begin{equation}
F^{\eta,h}_{\Gamma}:= \frac{-1}{\beta}\log \Tr \exp (-\beta H^{\eta,h}_{\Gamma}) \, ,
\end{equation}
 $\eta^{(\Lambda)}$ is the random field configuration obtained from $\eta$ by setting it to zero within $\Lambda$, and  $\condExp{\cdot}{\eta_\Lambda}$ is a conditional expectation, i.e. an average over the fields outside of $\Lambda$.
 (The modification of the field $h$ by $\pm\delta$ serves to select the desired ($m_\pm$) states).

 Somewhat inconveniently, it is not obvious that  for all models the limits in \eqref{gfDef}
exist.  Nevertheless, one  can prove that for each system of the class considered here there is a sequence of volumes $\Gamma_j \nearrow \Z^d$  for which the limit exists for all $\Lambda$, with convergence uniform in $\eta_\Lambda$.    The proof of this assertion is by a compactness argument, whose details can be found elsewhere~\cite{future}.

The essence of the proof of  Theorem 1 is the  contradiction between two estimates:
\begin{enumerate}[i.]
\item
Under Assumption~\ref{finiteRange}, equation~\eqref{finiteRangeIneq}:
\begin{equation}
|\G(\eta)| \leq  4 C | \partial \Lambda | \, . \label{uBound}
\end{equation}
\item 
Whenever $m_-< m_+$,
 $\G/\sqrt{|\Lambda|}$ converges in distribution  to a normal  random variable with a positive variance
   (as one would guess by considering the difference in the random field terms between states of  different mean  magnetizations,  neglecting the states' local adjustments to the random fields).
\end{enumerate}

More explicitly,
for the upper bound we note that in the absence of the interaction terms $V_\Lambda$, the right hand side of~\eqref{gfDef} would be zero.  Using~\eqref{finiteRangeIneq}, one gets \eqref{uBound}.

To prove the normal distribution for $\gf$, we apply a theorem of \citep[Proposition~6.1]{AW} (as corrected in \citep[p.~124]{bovier2006smd}).  It implies that for $\Lambda \nearrow \Zd$, under Assumption~\ref{fluct},  $\G/\sqrt{|\Lambda|}$ converges in distribution  to a normal  random variable with variance of the order of
\begin{equation}
b\ =\ \E{\frac{\partial\G}{\partial \eta_x}} = \epsilon ( m_+ - m_- ) \;.
\end{equation}

 The two statements described above contradict the assumption  that $m_-<  m_+$  in dimensions $d\le2$.   That is so even at the critical dimension,  where $L^{d/2} = L^{d-1}$.  The reason is that the lower bound
implies the existence of arbitrarily large fluctuations on that scale, whereas the upper bound is with a uniform constant.  This proves Theorem 1.\\

%

The above proof is similar to that of the classical results~\cite{AW,AW.PRL} which this work extends.  However, the discussion  of the free energy fluctuations was based there on the analysis of the Gibbs states, and more specifically of the response to the fluctuating fields of the `metastates' which were  specially constructed for that purpose.
Except for special cases, such as the QRFIM, that argument was not available for quantum systems, where  the equilibrium  expectation values are no longer given by integrals over positive measures.
 The proof of the quantum case is enabled by a more direct analysis of the free energy. \\

Theorem 2 is proven by establishing that in the presence of continuous symmetry the   upper bound  \eqref{uBound}, for $\Lambda = [-L,L]^d$,  can be replaced by:
\begin{equation}
|\G(\eta_\Lambda)| \leq K L^{d-2}\label{symUBound} \, .
\end{equation}
Here $\G$  is defined as in~\eqref{gfDef},\eqref{gLim}, but $\vec{h}=\vec{0}$, and $\delta$ is replaced by by $\vec{\delta} := \delta \hat{e}$ with $\hat{e}$ a unit vector.   This change in the upper bound raises  the critical dimension to $d=4$.

To obtain  \eqref{symUBound} we focus on 
\begin{equation}\label{grDef}
\gr(\vec{\eta}_\Lambda) := \condExp{F_\Gamma^{\vec{\eta},\delta \hat{e}} - F_\Gamma^{\vec{\eta},-\delta \hat{e}}}{\vec{\eta}_\Lambda} \, .
\end{equation}
Since
$
\G(\eta_\Lambda) = \lim_{\delta \to 0}\lim_{\Gamma \to \Z^d}
\frac{1}{2} \left( \gr(\vec{\eta}_\Lambda)  - \gr(0)    \right) \, ,
$
any uniform bound on  $|\gr|$ for given $\Lambda$ implies a similar bound on $|\G|$.  The claimed  bound may be obtained though a soft-mode deformation analysis, which we shall make explicit for the case of pair interaction (the general case can be treated by similar estimates).

The free energy $F_\Gamma^{\vec{\eta},-\delta \hat{e}}$ in \eqref{grDef} may be rewritten
by rotating both the spins and the field vectors  with respect to an axis perpendicular to $\hat{e}$ at the slowly varying  angles
\begin{equation}
\theta_x :=\ \left\{
\begin{array}{ll}
0, & \; \| x \| \leq L \\
\frac{\|x\|-L}{L}\pi, & L < \|x \| < 2L \\ \pi, & \|x\| \geq 2L
\end{array}
\right.  \, .
\end{equation}
The rotation aligns the external fields in the two terms ($\pm \delta \hat{e}$), except in the region $\|x < 2L\|$ where the effect is negligible when $\delta \to 0$.
The effect of the rotation on the random fields is absorbed by rotation invariance of the average.  In the end, the Hamiltonian of the rotated system differs from the Hamiltonian used to define the other free energy by
\begin{equation}\label{twist}
\begin{split}
\Delta H_{\underline{\theta}} := \sum_{\{x,y\} \subset \Gamma} & J_{x-y} \left[ \vec{\sigma}_x \cdot \vec{\sigma}_y \right. \\
- &\left. \vec{\sigma}_x \cdot \left( e^{i (\theta_y-\theta_x) \rho_y} \vec{\sigma}_y e^{- i (\theta_y-\theta_x) \rho_y} \right)
\right]
\end{split}\end{equation}

%

When the resulting expression for $F_\Gamma^{\vec{\eta},-\delta \hat{e}}$ in  \eqref{grDef} is expanded in powers of  $\theta_x-\theta_y \approx \pi \|x-y\|/L$,  the zeroth-order term cancels with $F_\Gamma^{\vec{\eta},-\delta \hat{e}}$, and the second and higher order terms yield the claimed bound.  The main difficulty is to eliminate the first order terms, which amount to a sum of $O(L^d)$ quantities each of order $1/L$.
However, the sign of these terms is reversed when the rotation is  in the reversed direction.
To take advantage of this, we combine two expressions for
$\gr(\vec{\eta}_\Lambda) $ with the rotations applied  in opposite directions, yielding:
\begin{equation}
\begin{split}
\gr(\vec{\eta}_\Lambda) &= \rm{Av}
\left[ \log \Tr e^{-\beta H} \right. \\
- \tfrac{1}{2} & \left. \log \Tr e^{-\beta (H + \Delta H_{\underline{\theta}})}
- \tfrac{1}{2}\log \Tr e^{-\beta (H + \Delta H_{-\underline{\theta}})} \middle|
\vec{\eta}_\Lambda \right]
\end{split}
\end{equation}
(where $H \equiv H_\Gamma^{\vec{\eta},\delta \hat{e}}$.)
By known operator inequalities \cite{Ruelle}:
\begin{equation}
\begin{split}
\log \Tr e^{-\beta H} - &\tfrac{1}{2} \log \Tr e^{-\beta (H + \Delta H_{\underline{\theta}})} - \tfrac{1}{2} \log \Tr e^{-\beta (H + \Delta H_{-\underline{\theta}})}
 \\ &\leq \tfrac{1}{2} \| \Delta H_{\underline{\theta}} + \Delta H_{-\underline{\theta}} \|
\end{split}
\end{equation}
The right hand side is zero to first order, and one is left with an \emph{upper bound} on~$\gr$ of the desired form.  Repeating this analysis with the roles of the terms exchanged we obtain an identical \emph{lower bound},  and thus  inequality~\eqref{symUBound} follows.\\


The  above argument is spelled out in detail in \cite{future}.   Let us end with few additional comments on the assumptions.
\newcounter{dcount}
\setcounter{dcount}{\value{ccount}}
\begin{list}{\arabic{ccount}. }{\usecounter{ccount} \setlength{\itemindent}{\leftmargin} \setlength{\listparindent}{\parindent} \setlength{\leftmargin}{0pt}}\setcounter{ccount}{\value{dcount}}
\item For Theorem~\ref{bigThm2}, the assumption that the interaction has a strictly finite range can be weakened to a condition somewhat similar to Assumption~\ref{finiteRange}.  For pair interactions (equation~\eqref{ONHam}) it suffices to assume: \begin{equation}
\sum_{x \in \Zd} |J_x| \ \| x \|^2 <\infty  \label{continuous_short} \, .
\end{equation}


 \item 
 The restriction to absolutely continuous distribution excludes a number of models of interest.  Such an assumption is generally necessary at zero temperature, as can be seen by the behavior of the Ising chain in a random field~\cite{Bleher} which takes only a finite number of values.  For positive temperatures it can be replaced by the requirement that the distribution has a continuous part which extends along the entire range of values.  For the QRFIM at finite temperature one need only assume that the random field has more than one possible value, and this may well be the case more generally.

\end{list}

\begin{acknowledgments}
We thank S. Chakravarty and M. Schechter for useful discussions.   M.A. wishes to thank the Weizmann Institute for the hospitality accorded him at the Department of Physics of Complex Systems.
The work of R.L.G. and J.L.L. was supported in part by NSF Grant DMR-044-2066 and AFOSR Grant AF-FA9550-04, and the work of M.A. by NSF Grant DMS-060-2360.
\end{acknowledgments}

\bibliography{gamma}
\end{document}